\documentclass[aps,pre,twocolumn,showpacs,epsfig,epsf,amssymb,floatfix]{revtex4-1}
\usepackage{graphicx}
\usepackage{amssymb}
\usepackage{amsmath}
\usepackage{subfigure}
\usepackage{booktabs}
\usepackage{float}
\usepackage{longtable}
\begin{document}
%\widetext

\title{Bacterial chromosome organization I: crucial role of release of topological constraints and  molecular crowders.}
\author{Tejal Agarwal$^{1}$ } \email{tejal.agarwal@students.iiserpune.ac.in}
\author{G.P. Manjunath $^2$}
\author{Farhat Habib$^3$}
\author{Apratim Chatterji$^{1,4}$} 
\email{apratim@iiserpune.ac.in}

\affiliation{
$^1$ IISER-Pune, Dr. Homi Bhaba Road,  Pune-411008, India.\\
$2$ Department of Biochemistry and Molecular Pharmacology, NYU Langone Medical Center, New York, NY 10016, USA.  \\
$3$ Inmobi - Cessna Business Park, Outer Ring Road, Bangalore-560103, India. \\
$^4$ Center for Energy Science, IISER-Pune,  Dr. Homi Bhaba Road,  Pune-411008, India.
}
\date{\today}
\begin{abstract}
We showed in our previous studies that just $3\%$ cross-links, at special points along the contour of the bacterial DNA  help the DNA-polymer to get organized at micron length scales \cite{jpcm,epl}.
 In this work, we investigate  how does the release of topological constraints help in the organization 
 of the DNA-polymer. Furthermore, we show that the chain compaction induced by the crowded environment in the bacterial cytoplasm contributes to the organization of the DNA-polymer. We model the DNA chain as a flexible 
bead-spring ring polymer, where each bead represents $1000$ base pairs. The specific positions of the cross-links have been taken from the experimental contact maps of the bacteria {\em C. crescentus} and {\em E. coli}. 
We introduce different extents of topological constraints in our model  by systematically changing the diameter of the monomer bead.  It varies from  the value where the chain crossing can occur freely  to the value where the chain crossing is disallowed.  We also study the role of molecular crowders by introducing an effective Lennard Jones attraction  between the monomers. Using Monte-Carlo simulations, we show that the release of topological constraints and the crowding environment play a crucial role  to obtain a unique organization of the polymer.

\end{abstract}
\keywords{}
%\pacs{87.15.ak,82.35.Pq,82.35.Lr}

\maketitle

\section{Introduction}
It is established  that the 3D organization of the DNA polymer (chromosome) in vivo in the bacterial or higher organisms is not random at large length scales ($>30$nm) \cite{le,aiden,benza,jun,tjong,marenduzzo,brocken,kremer,nicodemi,mirny,ganji,miriam,cagliero}. Thus, it is important to understand the factors which govern the organization of the chromosome, as the DNA-organization plays a crucial role in various cellular processes. In higher organisms, multiple chromosomes are packed  and confined inside the cell nucleus. In contrast, bacterial cells do not have a nucleus, and typically have just one or two chromosomes inside the cell. Thus, the bacterial chromosomes are not confined by the nucleus wall, and share the cell volume with other organelles/proteins/RNAs which affect the organization and dynamics of chromosomes in the bacteria \cite{jun,ralf,joyeux,debashish,martijn,xin,shane,dame}. It is  well known that the bacterial chromosomes occupy a significant fraction of the cell volume,  the section of the cell in which chromosome lies is called the nucleoid which is $\approx 15-25 \%$ of the cell volume for different bacterial cells \cite{joyeux}. All these conditions can affect and play a crucial role in the organization of the DNA polymer at large-length scales. In the last few decades many attempts have been made to understand the principles of  DNA organization in the bacteria as well as in higher organisms, especially, after the advent of new higher resolution experiments, e.g., Hi-C and single cell imaging techniques \cite{aiden,cagliero,qin,le,wang,youngren,ba,ganji}. 

The Hi-C experiments provide the contact map of the DNA segments for the bacteria (or higher organisms) at kilo-base (or mega-base) resolutions. The contact maps give the frequency of contacts between different DNA segments with each other at one kilo-base resolution. The contact map data provided by these experiments are the average over cell populations. These experiments give important insights about the structure of the chromosome at large length scales but the physical modeling of the chromosome is necessary to conclusively determine the 3D spatial organization as well as the factors which  govern the organization. 

To understand the organization of the chromosome at large length scales much above the $30 nm$ fiber researchers use polymer physics principles where the chromosome is modeled as a coarse-grained bead-spring polymer \cite{le,ralf,elcock,amos,geoffrey,marenduzzo,jun,ganai,mirny,li,nicodemi} and incorporate other interactions which is expected to be relevant for chromosome organization. Bacterial chromosomes are typically ring polymers, whereas DNA chains of higher organisms are linear polymers. Researchers model the DNA-binding-proteins by the additional diffusing  particles which bind together two different DNA segments from two different part of the chain \cite{geoffrey,marenduzzo,nicodemi}. In one such model the proteins can diffuse along the polymer chain contour (loop-extrusion model) \cite{geoffrey} and in the other model the spherical particles can diffuse in the 3D space and bind or unbind two DNA segments (switch and binders model) \cite{nicodemi}. 

Other researchers have also looked into the effect of molecular crowders in the organization of the chromosome but neglect the effects of DNA binding proteins \cite{ralf}. It is shown in the previous theoretical and experimental studies that the DNA polymer compacts and  collapses in the presence of the crowding environment  (DNA condensation) \cite{ralf,kojima}. In the presence of the charged molecular crowders the effective interactions between DNA-DNA segments increases which leads to the segregative phase separation \cite{murphy,ralf,joyeux}. The segregative phase separation may happen if two crowders (e.g., protein, RNAs, etc.) or a crowder and DNA segments of the same charge repels each other more strongly than two different segments of the DNA \cite{joyeux}.  Another simulation study of ring polymers with uncharged molecular crowders and  cylindrical confinement show that increasing the density of the crowding molecules leads to the decrease in the radius of gyration $R_g$ of the ring polymer due to the depletion interaction between the polymer and the molecular crowder \cite{ralf}. In that study, the monomer-monomer, crowder-crowder, and polymer-crowder interactions had been modeled by purely repulsive LJ potential.  Other studies which incorporate dissimilar crowders also established that the crowding environment helps in the compaction of the DNA molecule in the nucleoid region inside the cell \cite{elcock}. 

In addition to these molecular crowders,  bacterial cells also have the enzyme topoisomerase which helps the DNA chain to overcome topological constraints by suitably cutting and rejoining the chain and hence allowing the chain to cross itself. Researchers model the activity of enzyme topoisomerase explicitly by taking a suitable potential \cite{le} to allow chain crossing or by taking the  diameter of the monomer beads smaller compared to the bond length between adjacent monomers \cite{jpcm,epl}. More detailed description of chromosome also incorporates explicit  plectonemes (the loopy patterns formed by the DNA super-coiling) modeling \cite{le,william}, where the length and size of the monomers constituting plectonemes are  chosen as a parameter. Using the multi-parameter optimization researchers try to get the best set of parameters which fits well with the experimental DNA contact map. 

The research on the bacterial DNA polymers are not limited to studies of the organization  at the stage of the cell cycle when the cell is not dividing, but several attempts have also been made to understand the dynamics of chromosomes under cylindrical confinement at different stages of the cell cycle \cite{jun}. It has been shown that the cylindrical confinement helps in the segregation as well as in the organization of two chromosomes after replication (Mother and daughter chromosome) \cite{jun2,byha}. For other studies which focus on the polymer under confinement see \cite{milchev,dai,wang,austin}. 

\subsection{Our previous work}

Previously, we have  investigated the organization of the bacterial chromosome, where we focused on the role of cross-links (CLs) at very specific positions along the chain contour and their effect on  the organization of the chromosome of bacteria {\em C. crescentus} and {\em E. coli} \cite{jpcm,epl}. We modeled the chromosome as a bead-spring ring polymer. The DNA of {\em E. coli} has $4642$ kilobase pairs and the DNA of {C. crescentus} has $4017$ kilobase pairs. We modeled the DNAs of {\em E. coli} and {C. crescentus} as ring polymers with $4642$ and $4017$ monomers respectively. We took a relatively small value of the monomer diameter, i.e., $\sigma=0.2a$, (where the bond length $a$ was set as $a=1$) to allow the chain to cross itself. The positions of the cross-links were chosen from the experimental contact maps (obtained from Hi-C experiments) of bacterial DNA {\em E. coli} and {\em C. crescentus} \cite{le,cagliero}. We cross-linked the specific pair of monomers which have contact map  probabilities higher than a certain threshold; we can set different values of the threshold to have different numbers of cross-links in our chain.  We termed these different sets of cross-links at specific positions along the chain as biological cross-links BC-1 and BC-2; BC-1 set has fewer CLs than BC-2 and BC-1 is a subset of BC-2. Two different segments of the chain (which need not be neighbors along the chain contour) can be held spatially close to each other, i.e., cross-linked, due to the presence of  DNA binding proteins. The  BC-1 and BC-2 sets of CLs for {\em E. coli} and {C. crescentus} are different. 

Using Monte-Carlo simulations, we established that the BC-2 set of cross-links at these  positions ($\approx3\%$ of the monomers) give rise to a particular well-defined structure of the polymer for both  bacteria. There were effectively $82$ and $60$ independent CLs for {\em E. coli} and {C. crescentus}, respectively. We had also shown that if we randomly chose an equal number of monomers as in BC-1 and BC-2 and cross-link them, (the set of CLs at random positions are termed as RC-1 and RC-2), the nature of the organization of the polymer is different from the polymer with biological cross-links.  Using the statistical quantities, we predicted the overall 2D organization of the polymer with inputs from chromosomal contacts maps of bacteria {\em E. coli} and {C. crescentus}, respectively. We also compared and validated our predicted organization of the polymers having BC-2 CL-sets with the experimental data of bacteria {\em E. coli} and {\em C. crescentus}. We did not take into account the effect of molecular crowders and cell confinement in the organization of the chromosome in our previous study.

\subsection{Scope of our present investigations}

Motivated by the success of our previous studies, we now want to understand the role of other factors in the organization of the DNA polymer with the CLs at specific positions. In particular, we want to investigate the following aspects: (a) how does the organization of the DNA get affected if the DNA polymer cannot overcome the topological constraints by crossing  itself?  We can decrease the probability of chain crossing  by increasing the bead diameter of the monomers such that chain crossing becomes more difficult. (b) How does the DNA polymer organization in bacteria change in the crowded environment in the bacterial cytoplasm? We can introduce the effect of crowders by introducing a weak Lennard Jones attraction between the monomers. The results will also be equivalent to the investigation of the effect of changing solvent quality around the DNA-polymer. (c) Lastly, what is the effect of the cylindrical confinement of the cell walls in the organization of the bacterial DNA polymer?

As mentioned before, to control the relative ease of chain crossing, we systematically vary the diameter of monomers which in turn affects the excluded volume (EV) interactions between monomers. It is non-trivial to a priori understand the effect of increasing EV on the organization of the cross-linked DNA-polymer. On the one hand, it has been reported for bottle-brush polymers that the persistence length of the primary chain increases with increase in the density of monomers  in the side-chains.  In particular EV interactions in dense polymers with  branched or side chain architecture could promote particular architectures \cite{hsu_binder,debashish}. In our case, we can think that increasing EV would lead to less availability of phase space for the monomers, especially at the core of the globule, where the density of cross-links is higher. This can help the polymer to get organized in which each segment of the polymer will have well-defined neighboring segments. But, increase in EV interactions will increase the topological constraints and the polymer may not be able to relax in a particular  organization. So we did a systematic study on the effect increase of the diameter $\sigma$ on the organization of the polymer with special CLs. 

To be able to claim  that a polymer  gets organized  due to the presence of CLs, we  started our Metropolis MC run with $9$  different initial configurations of the polymer, and calculate ensemble averaged structural quantities to quantify if the polymer reaches the same organization for all the $9$ independent runs \cite{jpcm,epl}. We are well aware that the live bacterial cells are non-equilibrium systems, but we assumed the chromosome to be in a state of local equilibrium, especially at the stage of the cell cycle when the cell is not dividing. We maintain very similar approaches and assumptions in our current investigation.

The manuscript is organized as follows. The next section describes
the methods and modeling in greater detail. This is very similar to  what is described in our previous communications, 
except that we now study the effects of (a) increasing bead size which affects the frequency of chain crossing (b) effective attractive interaction between monomers induced by the crowders. The consequence of changing these factors are described in section III, the Results section. The Results section is divided into two subsections where we focus on the effect of systematically varying one of the above-mentioned factors in each of the sub-sections. We give comprehensive data for our analysis of {\em C. crescentus} but only conclusive quantitative results for {\em E.coli}. Our conclusions derived from the study of both model-chromosomes remain nearly identical, which reassures us about the accuracy of our results. We finally conclude the manuscript with Discussions. 

The investigation on the role of confinement is more involved, and its effect on the organization of the chromosome is seen to be crucial. Thereby the role of confinement is described in accompanying separate manuscript which constitutes Part-II of our study. In that study, we also comment on the combined effects  of confinement as well as the role of crowder molecules which induce effective attraction between DNA-segments.

\section{Model and simulation method.}

We model the circular chromosome of the bacteria {\em E. coli}  and {\em C. crescentus} as a flexible bead-spring ring polymer having 
$N=4642$ and $N=4017$ monomers, respectively, with cross-links at specific positions chosen from the experimental contact map of bacterial DNA as in our previous work \cite{jpcm,epl}. There, we have already enumerated the list of monomer pairs which are cross-linked, we termed them as BC-1 and BC-2. The bacteria {\em E. coli} and {\em C. crescentus} each consist of $\approx 4$ million base-pairs. Thus, our coarse-grained monomer represents $1000$ base-pairs of (BP) the DNA-polymer similar to the resolution of the Hi-C experimental contact map. Starting from very different initial conditions, we let the polymer chain equilibrate with the cross-links at specific positions, details of this can be found in \cite{jpcm,epl}. We generate different  micro-states of the system using Monte-Carlo simulations to calculate the average statistical quantities of interest. We then compare and analyze the different structural quantities from independent runs, from which we can comment whether the chromosome structure remains
robust (or not) across independent runs with the change in the system parameters.

We now describe our model of the polymer with cross-links. The nearest neighboring monomers along the chain contour interact by a harmonic potential $V=\dfrac{1}{2}\kappa(r-a)^2$;
where $a=1$ is the mean bond length, and we choose $a$ as the unit of length for our simulations. We fix the spring constant to be $\kappa=200k_{B}T/a^2$  and $r$ is the distance between the monomers at a particular micro-state. Thermal energy $k_{B}T$ is the unit of energy and we take $k_{B}T =1$. The excluded volume interaction between the monomers is modeled by the Lennard Jones potential $V_{LJ}(r) = 4 \epsilon \left[ (\sigma/r)^{12} - (\sigma/r)^6 \right]$ truncated at  $r=2^{1/6}\sigma$ and suitably shifted to zero (Weeks Chandler Anderson potential) with $\epsilon=1 k_BT$. However, when 
we introduce a weak attraction, the cut-off is set at $R=3 \sigma$ for different values of $\epsilon$. We bind (cross-link) the specific monomers (CLs), in the polymer which are found in spatial proximity with high probability in the experimental contact maps of bacteria {\em E. coli} and {\em C. crescentus}. The cross-linked monomers interact with harmonic potential $V_{c}=\dfrac{1}{2}\kappa_c(r_{cc}-a)^2$, we take $\kappa_c=200k_{B}T/a^2$ same as the spring constant between the nearest neighbor monomers along the contour. Here $r_{cc}$ is the distance between the cross-linked monomer pair. 

In our previous study, we took two CL sets with $49$ and $153$ number of CLs and termed as BC-1 and BC-2, respectively for the bacteria {\em C. crescentus}. The number of CLs for the bacteria {\em E. coli} has been taken to be $47$ and $159$ corresponding to the BC-1 and BC-2 CL sets.
The number of CLs in each set was taken by setting different frequency cutoff in the experimental contact map of the bacteria {\em E. coli} and {\em C. crescentus}.
Thus, BC-1 CL set is the subset of the BC-2 CL set. To quantify the organization of a polymer chain we measured large-length scale positional correlations between the center of mass (CM) of different segments of the chain;
each segment consisted of $50$ (or $58$ monomers for {\em E. coli}) monomers such that the entire ring polymer consists of $80$ segments. In case of biological cross-links we observe that some CLs were not independent of the others, e.g. if, monomer i and j are found in proximity in the experimental contact map, and if monomer i+1 is close to the monomer j+1 then these should not be considered as two independent CLs, but they are effectively one cross-link. Thus for the bacteria {\em C. crescentus} ({\em E. coli}) there are {\em effectively} $26$ and $60$ ($26$ and $82$) CLs instead of  $49$ and $153$ ($47$ and $159$) CLs, respectively. 
We will use the same terms, BC-1 \& BC-2 sets to denote $49$ \& $153$ number of CLs (effectively $26$ \& $60$) chosen from the experimental contact map of the bacteria {\em C. crescentus}, unless otherwise mentioned. To study the effect
of chain crossing, we systematically change the value of $\sigma$ for a polymer with the BC-2 set of CLs and measure different statistical quantities to identify the structure of the polymer.

As mentioned in the introduction, we also study DNA-polymer organization in the presence of the BC-2 set of cross-links as we introduce an attraction between the monomers. The attraction between the monomers mimics the role of molecular crowder.
We keep all parameters and constraints the same as described earlier, the only change is in the increase in the range of cutoff of the LJ potential and the corresponding value of $\epsilon$. The monomers which are not neighbors in the chain interact by the LJ potential $V=4\epsilon\left[(\sigma/r)^{12}-(\sigma/r)^6\right]$ suitably cut at $r_c=3.0 \sigma$ and shifted. By tuning the value of parameter $\epsilon$ in the potential we can have different values of attraction strength between the monomers. We do not want to investigate the collapse of the DNA-polymer with changing solvent conditions. Thus, we restrict ourselves to  values of $\epsilon/k_BT < 1$, such that the local density of monomers is less than that observed for a collapsed globule state and the monomers 
can still explore phase space without getting kinetically trapped. We fix the monomer's diameter $\sigma=0.2a$ while we change the value of the parameter $\epsilon$ of LJ potential. 

For both the studies of the parameters $\sigma$ and $\epsilon$, we start our simulation from $9$ independent initial conditions and check the polymer organization across configurations obtained after equilibration from these initial conditions. 
%We have designed the $9$ initial independent condition such that the
%CL monomers are at very different positions with respect to each
%other in every initial condition at the beginning 
%of the simulation. 
In all $9$ different initial conditions, the distance between the adjacent monomers along  the ring-polymer contour is maintained at a distance of $a$, but the monomers which constitute the CLs  do not maintain any such precondition, and the distance  between them is decided by the geometry of the initial configuration, for details about the initial conditions refer \cite{jpcm}. However, as the chain reaches equilibrium, the cross-linked monomers do come near each other and maintain an average distance of $a$. 

We start the simulations from the initial monomer configurations with a very small value of the spring constant  between cross-links, $\kappa_{c} =0.2k_{B}T/a^2$, and after every $1000$ MCS we increase the value of $\kappa_c$ by $0.2k_{b}T/a^2$.  Thus, at the end of equilibration after $10^6$ MCS the spring constant will have a value  $\kappa_{c} =200k_{b}T/a^2$. We check for equilibration by checking that the average  energy per monomer is the same across the $9$ independent runs. Moreover, we explicitly checked that none of the bonds are stretched more than $5\%$ of its equilibrium length,  which is possible if the absence of 
chain crossing prevents the polymer from relaxing to equilibrium configurations due  to topological constraints.  After this equilibration, we simulate for $1.2\times10^{7}$ MCS and start collecting data after every $5$ MCS to calculate the average statistical quantities. Moreover, for the above-mentioned studies where we vary either $\sigma$ or $\epsilon$, we  use a very large simulation box ($200\times200\times200 a^3$) with no confinement. 
We chose the large box size to avoid interactions between the polymer due to the periodic boundary condition. 

\section{Results}

Before we present  how  (a)  changing the excluded volume parameter $\sigma$ (b) changing the value of $\epsilon$ affects the organization of the chromosome,  we first present the analysis of how the increase in monomer diameter $\sigma$ changes the size of the polymer.  This and other statistical quantities that we first discuss  will help 
us in understanding and analyze our subsequent results. 

\subsection{ Effect of varying the value of $\sigma$.}

For Gaussian linear polymer chains (without considering excluded volume (EV)  interactions between monomers), the radius of gyration $R_g \sim N^{0.5}$, whereas if the polymer is self-avoiding then $R_g \sim N^{0.6}$. For the self-avoiding ring polymers $R_g \sim N^{0.65}$ \cite{bishop}. Self-avoidance can be modeled by a bead-spring model of a polymer chain with EV interactions modeled by a suitably truncated Lennard Jones potential between beads to retain only the repulsive part of the potential. It has been shown for linear polymers that any value of $\sigma/a \neq 0$ gives the value of scaling exponent to be $0.6$ though the absolute value of $R_g$ decreases \cite{binder}.
An estimate of the size of the polymer is given by the radius of gyration  $R_g$ of the chains. The radius of gyration is calculated as $\sqrt{(I_1 + I_2 + I_3)/2M}$, where $I_1,I_2,I_3$ are the eigenvalues of the moment of inertia matrix and $M=mN$; we have taken $m=1$.   For our studies, we have a ring polymer  with cross-links at very specific points along the chain; we cannot expect the scaling which is relevant for linear or a ring polymer without CLs. For our studies,  we focus on the increase in the size of the DNA ring-polymer of fixed length and specific CLs with the increase in the EV parameter $\sigma$.

\begin{figure}[!hbt]
\includegraphics[width=0.6\columnwidth]{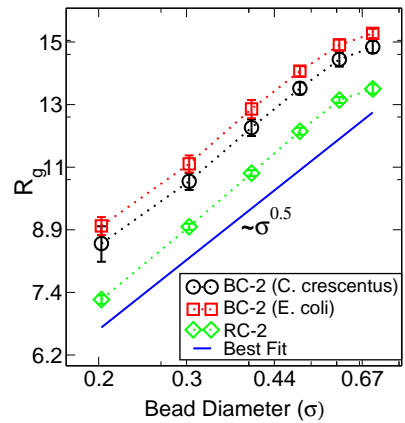}
\caption{\label{fig1}
The plot shows the value of the radius of gyration ($R_g$) in a log-log plot for different values of bead diameter $\sigma$. Different lines correspond to the $R_g$ of polymer with BC-2 sets of CLs chosen from the experimental contact map of {\em C. crescentus}, {\em E. coli} and with the RC-2 set of CLs. The average is taken over $9$ independent initial conditions and the standard deviation is shown by the error bars. 
}
\end{figure}

To this end, we plot $R_g$ versus $\sigma$ in Fig. \ref{fig1}(a) for the polymer with BC-2 and RC-2 CLs in a log-log plot. 
The error bars in Fig. \ref{fig1}(a) show the standard deviation in the value of $R_g$ for the polymer starting from $9$ independent initial conditions.  From Fig. \ref{fig1}(a) we see that the value of $R_g$ increases for polymer with the BC-2 and RC-2 from $9a$ to $15a$ and $7a$ to $13a$, respectively. The slope of the graph gives the exponent for the scaling relation. From the graph we see that the value of $R_g$ scales as $R_g \sim \sigma^{0.5}$ for the polymers with BC-2 and RC-2 CLs.  

To analyze  this, we note that in our previous studies we observed that with the BC-2 set of CLs, the CL monomers had clustered towards the center of the coil and there were lengthier loops of monomers which were in the peripheral region of the globule \cite{jpcm,epl}. On the other hand, for the RC-2 CL set, the CL monomers were distributed along the contour and  in space randomly. As a consequence, there were not relatively longer loops present on the periphery of the polymer globule. Due to this, there was higher compaction of the globule and the value of $R_g$ in case of polymer with RC-2 CLs was less compared to the  polymer with BC-2 CLs. Now the question is, with the increase in the value of  $\sigma$ will the increase in $R_g$ is due to the overall swelling of the chain with BC-2 CLs, or is the amount of swelling is affected by the location of CL monomers in the chain contour? Since the exponent in Fig. \ref{fig1} remains the same for both RC-2 and BC-2 set of CLs, it appears that there is an overall swelling of the chain. We conclude this because the increase in the size of the polymer does not depend on the locations of the CL monomers in the chain contour.

The value of $R_g$  gives only an estimate of  the overall extent of the polymer but does not give the information about the internal reorganization of monomers due to the increase in $\sigma$. In the case of polymer with BC-2 CL set, the clusters of cross-links pull a large number of other monomers near the center of the coil and increase the monomer density in the inner part of the globule. To check how the radial distribution of monomers changes with the  parameter $\sigma$  we calculate the number density of monomers from the center of mass (CM) of the polymer globule.    The next question is  whether relatively taut stretches of polymer  between two individual CLs hold these clusters of CLs near the center of the coil? In this scenario, if $\sigma$ is increased, the average distance between CL clusters and their spatial locations could remain relatively unaffected, and the effect of swelling could be significant only at the peripheral loops.   In the other scenario, the core region will swell along with the peripheral regions of the polymer coil.
\begin{figure}[!hbt]
\includegraphics[width=0.49\columnwidth]{no_density_exvol.eps}
\includegraphics[width=0.49\columnwidth]{cumulative_c_no_density_exvol_caul.eps} \\
\includegraphics[width=0.85\columnwidth]{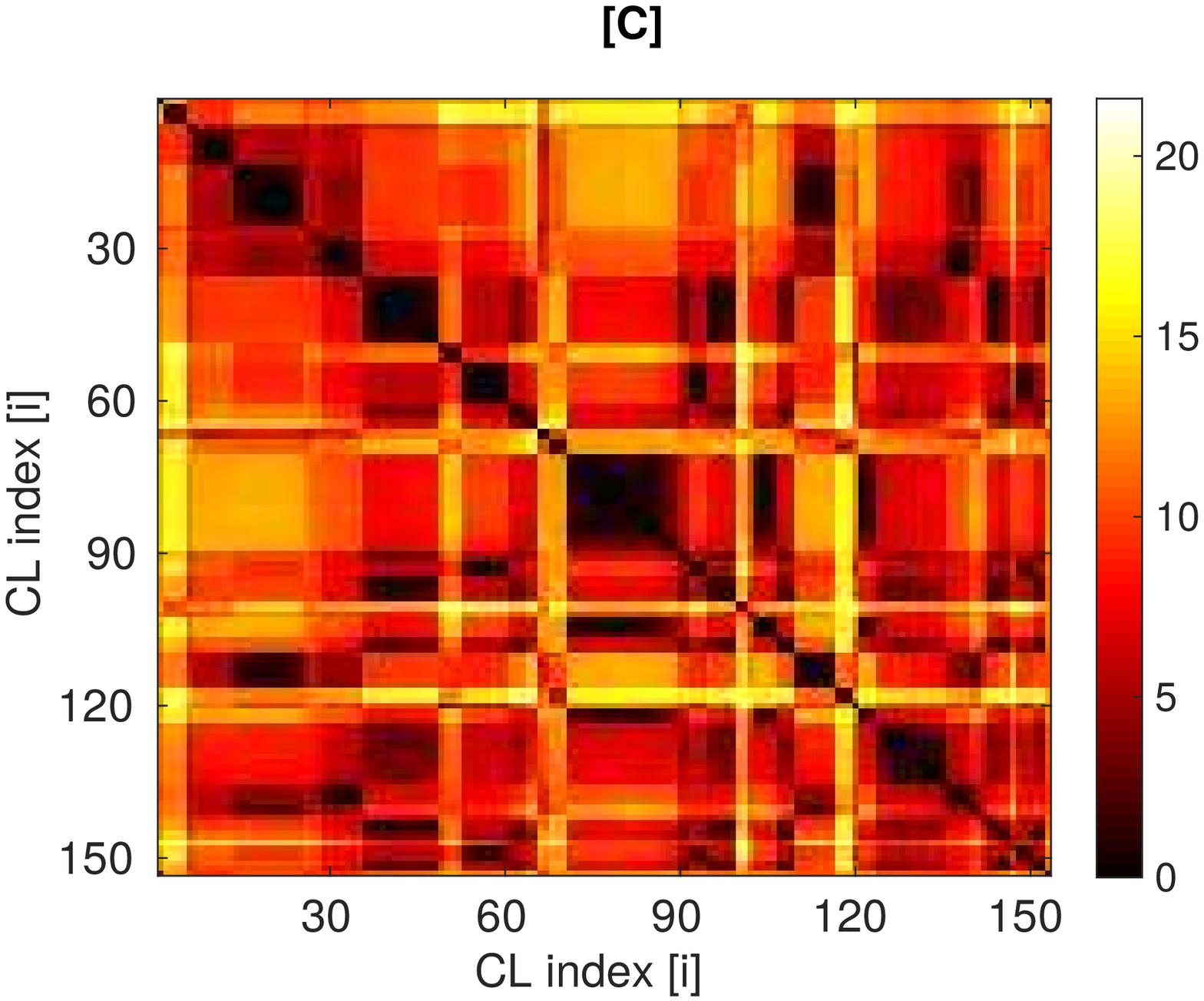} \\
\includegraphics[width=0.85\columnwidth]{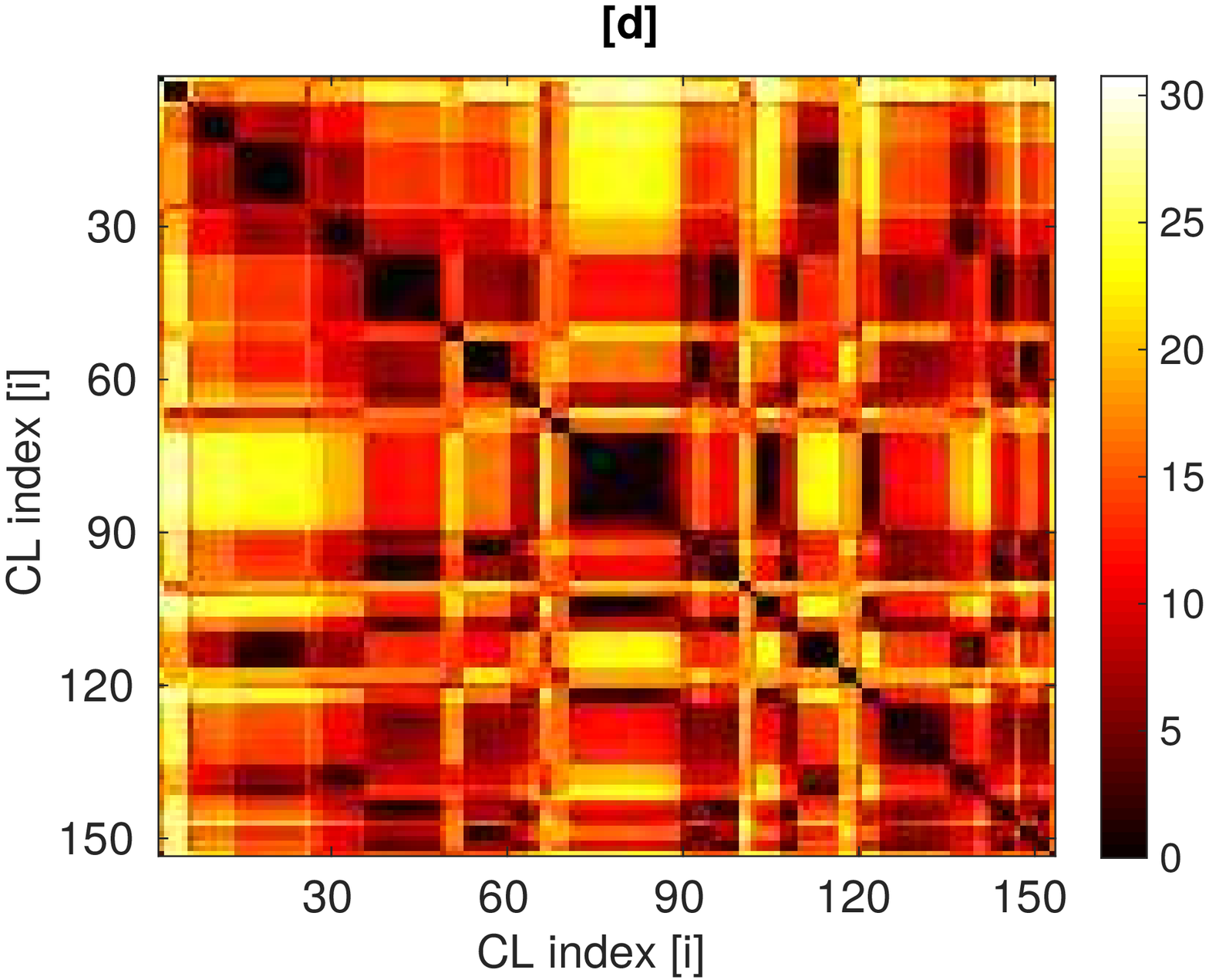}
\caption{\label{fig23}
(a) The plot shows the number density of monomers $n_{M}(r)$ as a function of the
distance (r) from the center of mass of the globule for different values of parameter $\sigma$. (b) The graph shows the cumulative number of CLs normalized by total number of CLs as a function of distance from the CM of the polymer for different values of bead diameter $\sigma$. The colormaps (c) and (d) represent the average distance in units of $a$ between the 
CL monomers i and j for $\sigma = 0.3a$ and $0.7a$, respectively. The color denotes the average distance 
between CLs $i$ and $j$. The graphs correspond to the model chromosome with the BC-2 set of CLs of bacteria {\em C. crescentus}.
}
\end{figure}

To investigate this, in Fig. \ref{fig23}(a),(b) we plot the number density of the monomers $n_{M}(r)$ and the cumulative number of CLs $c$-$n^{*}_{M}(r)$, normalized by total number of CLs, respectively,   as a function of the radial distance $r$ from the center of mass (CM) of the globule. The quantities   $n_{M}(r)$ and $c$-$n^{*}_{M}(r)$ are plotted for different values of the parameter $\sigma$. 
The error bars show the standard deviation from the average value in $9$ independent initial conditions.
From the Fig. \ref{fig23}(a), it can be confirmed that as we increase the value of  parameter $\sigma$ the value of $n_{M}(r)$ decreases near the center of the coil (for low values of $r$). This indicates that there is a swelling of the polymer core with the increase in the value of $\sigma$. But $n_{M}(r)$ has non-zero values even at larger $r$ when the  the value of $\sigma$ is increased.
The normalized cumulative number of CLs, $c$-$n^{*}_{M}(r)$ is shown in the Fig. \ref{fig23}(b). From the figure, we see that for low values of $\sigma$ (say $\sigma=0.2a$) the value of $c$-$n^{*}_{M}(r)$ from the CM of the chain. 
Comparatively,  $c$-$n^{*}_{M}(r)$ reaches $1$ at $r=15a$ for $\sigma=0.5a$. These results again indicate that the increase in the size of the polymer is because of the overall swelling of the chain as the number of monomers, as well as the number of CL monomers, increase for larger values of r for larger $\sigma$. 

We also check for the average distance between the CL monomers for different values of parameter $\sigma$. This is shown in the Figs. \ref{fig23}(c),(d) as the colormaps. The x and y-axis represent the CL index, and the color represents the average distance between the monomers constituting the CL, note that the range of colorbars is different.  The dark color represents the two CLs are in proximity, and the bright color represents the two CLs are far from each other. The average value of the distance is given in the colorbar.
The two representative  colormaps in the figure correspond to $\sigma=0.3a$ and $\sigma=0.7a$, respectively. From the figure, we find that the average distance between the CL monomers increases with $\sigma$, but the pattern of dark and bright pixels in the colormap remains the same. This suggests 
that there is an overall swelling of the CL cluster with the increase in the value of EV parameter $\sigma$ hence the increase in the value of $R_g$. 

\begin{figure}[!hbt]
\includegraphics[width=0.75\columnwidth]{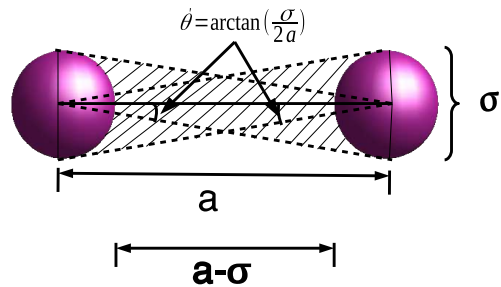}
\caption{\label{fig14}
The schematic diagram shows the shaded region between the two neighboring monomers (along the chain contour) where the third (non-neighboring) monomer can come.
We calculate the average number of monomers whose centers lie in the shaded region to calculate $\langle n_{CC} \rangle$. For details refer to the text.
}
\end{figure}

We expect that the crossing of chains should not be feasible once the value of the parameter $\sigma >0.4a$. Even if we introduce  Monte-Carlo attempts with trial large-displacement of monomers to encourage chain crossing,  chain crossings will not occur as  large trial displacements MC attempts leads to a high energy cost arising from large extensions of harmonic bonds. For smaller values of $\sigma$, the chain crossings can occur  frequently and the probability $P_{CC}$ of chain-crosses should decrease as we increase the value of $\sigma$. It is difficult to calculate the exact frequency of chain crossing in our simulations. So, to estimate the frequency of chain crossing in each Monte-Carlo step for different
values of EV parameter $\sigma$ we calculate the quantity $\langle n_{CC} \rangle$, which gives the average number of monomers (from the other parts of the chain) between two neighboring monomers along the chain. It is a reasonable estimate because  monomers which are in between two adjacent monomers along the chain can either cross the chain in next MC attempt or move away.

\begin{figure}[!hbt]
\includegraphics[width=0.6\columnwidth]{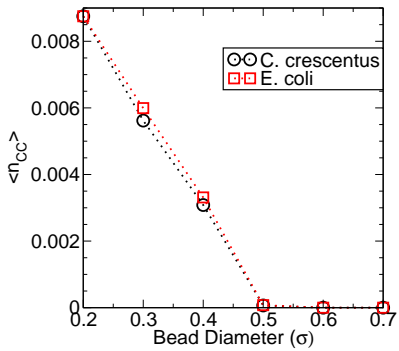}
\caption{\label{fig22}
The plot shows the Bead diameter $\sigma$ on the x-axis, and the quantity $\langle n_{CC} \rangle$ which is an estimate of 
chain crossing is plotted 
on the y-axis.  Refer text for the precise definition of the quantity $\langle n_{CC} \rangle$. 
}
\end{figure}

We calculate the quantity $\langle n_{CC} \rangle$ as follows. The distance between the nearest surfaces of two neighboring monomers  along the chain $i$ and $i+1$ is $a-\sigma$, refer to the schematic diagram of Fig. \ref{fig14}. 
Suppose another monomer j comes in between the two monomers $i$ and $i+1$.  The angle $\theta^{'}$ between the vector joining the center of the monomer $i$ and $j$ (or alternatively between $i+1$ and $j$) and the vector joining the center of the monomers $i$ and $i+1$ should be $\theta^{'} \le tan^{-1}(\frac {\sigma}{2a})$. It is shown as shaded region in the schematic diagram of Fig. \ref{fig14}. 
We calculate the number of monomers which are satisfying the above condition and normalize it by the total number of monomers $N$ to obtain $\langle n_{CC} \rangle$. The quantity $\langle n_{CC} \rangle$ is plotted in the Fig. \ref{fig22} versus the EV parameter $\sigma$. It is  averaged  over $9$ MC runs starting from $9$ independent initial conditions.  From the figure, we see that on increasing the value of the parameter $\sigma$ the value of $\langle n_{CC} \rangle$ decreases and for $\sigma \ge 0.5$ the value of $\langle n_{CC} \rangle$ becomes zero, as expected. From this data, we can estimate that the frequency of chain crossing drops rapidly with increase in the value of $\sigma$. 

Now with this background,  we would like to focus on how  increasing $\sigma$ affects the overall {\em organization} of the DNA-polymer. Instead of calculating radial distribution functions $g(r)$ of different segments of the chain, we calculate  the positions of different chain segments with respect to each other to quantify and analyze the organization of a chain. For this, we calculate the positional correlation between different segments as we did in our previous studies \cite{jpcm,epl}.
We determine the position of the center of mass (CM) of each segment (50 monomers each) and calculate the probability of the CM of two segments  to be at a distance less than a cutoff distance, $R_c$. In our previous papers, we chose the value of $R_c=5a$ to be nearly half of the value of $R_g=9a$ for $\sigma=0.2a$. Since  $R_g$ scales as $R_g\sim \sigma^{0.5}$ with different values of the EV parameter $\sigma$, we choose the values of $R_c$ for this study, such that the ratio $R_c/R_g\approx 0.55$ is maintained consistent with our  previous studies \cite{jpcm,epl}.  
(Note that we incorrectly calculated the value of $R_g$ to be $\approx 7a$ for $\sigma=0.2a$ in \cite{epl},
as we used the expression $R_g=\sqrt{(I_1+I_2+I_3/3M)}$. The correct formula is  $R_g=\sqrt{(I_1+I_2+I_3/2M)}$.
Using this correct expression, $R_g \approx 9a$.) 

\begin{figure}[!hbt]
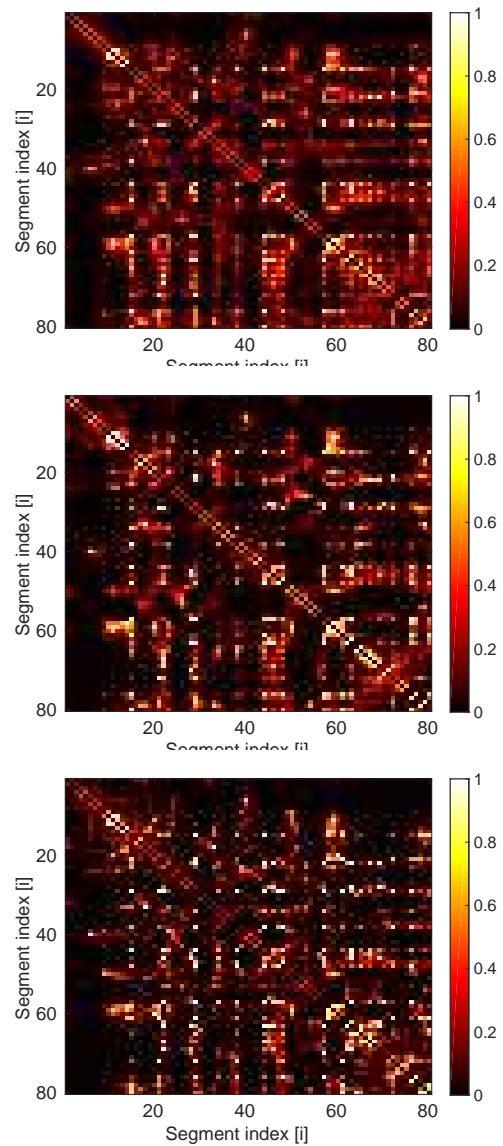

\includegraphics[width=0.75\columnwidth]{caulCL_cm_dom_corr_exvol02_eps0.eps} \\
\includegraphics[width=0.75\columnwidth]{caulCL_cm_dom_corr_exvol04_eps0.eps} \\
\includegraphics[width=0.75\columnwidth]{caulCL_cm_dom_corr_exvol07_eps0.eps}
\caption{\label{fig16}
The plots show the positional correlation colormaps of different polymer segments for three value of bead diameter $\sigma$ for the model chromosome of bacteria {\em C. crescentus}. The upper colormap is for $\sigma=0.2a$, middle is for $\sigma=0.4a$ and the lower colormap correspond to $\sigma=0.7a$.  We present only one colormap for each value of $\sigma$, though we have $9$ other 
colormaps from independent runs for each value of $\sigma$. While calculating the positional correlations we keep the value of $R_c/R_g=0.55$. Refer, the text for details.
}
\end{figure}

The positional correlations between the CMs of different segments are shown in colormaps of Fig. \ref{fig16} for $\sigma=0.2a, 0.4a, 0.7a$, respectively. In the colormaps, the x-axis  and the y-axis
represent the segment index corresponding to the CM of the $80$ segments. The color represents the probability
of the two segments to be within a cutoff distance $R_c$. Bright color shows the higher probability
of two segments to be within cutoff distance $R_c$.  From Figs.\ref{fig16} we see that as we increase the value of the parameter $\sigma$ from $0.2a$ to $0.7a$ the rectangular patch like-pattern of the colormaps start disappearing and for $\sigma=0.7a$ the rectangular patch-like pattern completely disappears. The rectangular patch like-pattern (where all the pixels are near of the same color)
in the colormaps show that the neighboring segments  along the chain contour of a particular segment-$i$ (50 monomers) come close in the 3D space with the same probability to another segment $j$, where $i$ and $j$ are far 
apart along the contour. Also, the number of bright pixels in the colormaps  decreases in-spite of choosing the higher values of $R_c$ with increasing $R_g$ in each case. This has also been quantified later in the result section. Hence, there are indications to conclude that with the increase in the value of  
$\sigma$, there is a loss of organization or structure of the DNA-polymer. We also get similar qualitative results for the colormaps of the polymer with the BC-2 set of cross-links chosen from the experimental contact map of bacteria {\em E. coli}.

\begin{figure}[!hbt]
\includegraphics[width=0.6\columnwidth]{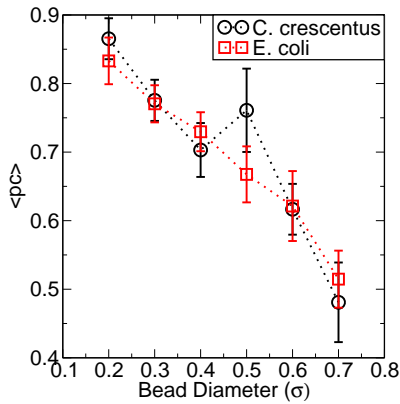}
\caption{\label{fig3}
The y-axis shows the average value of the Pearson correlation ($\langle pc \rangle$) of positional correlation colormaps with parameter $\sigma$ on the x-axis. The average is taken over ${}^{9}C_{2}=36$ values of Pearson correlation. Error bars represent the SD from the mean value. The correlation is calculated over only those pixels ($i,j$) for which 
$p_{ij}>0.05$ in at least one of the runs. This is done to prevent large contribution to $\langle pc \rangle$ from the regions which are dark in all color maps. 
}
\end{figure}

Till now, we have been identifying the organization from the visual inspection of the color-maps, but we need more suitable quantities to quantify and compare the level of organization. We can claim that a particular ring polymer gets organized in the presence of CLs only if the colormaps from different MC runs (starting from different initial conditions) relax to the same structure. As seen previously the BC-2 CLs lead a unique organization/structure of the polymer, and all the nine independent runs give the same positional correlations colormap for $\sigma=0.2a$. With the increase of the value $\sigma$, this will remain true unless the polymer gets kinetically stuck in different configuration.
To quantify the similarity of colormaps, we calculate the Pearson  correlation between the colormaps of positional correlations from the runs starting from $9$ independent initial conditions. A high value of  Pearson correlation on comparing positional correlation colormaps from different runs is indicative of the same resultant organization  of the polymer chain due to CLs. This, in turn, will lead to a statistical measure which will quantify how well-defined  a structure is, in spite of the presence of thermal fluctuations.

For a particular value of $\sigma$, the $9$ independent runs give ${}^{9}C_{2}=36$ comparisons between pairs of colormaps, and thereby $36$ values of Pearson correlation. We can calculate the average value of pc, and also get the SD (standard deviation) from the mean. Low values of pc and large SD values represents positional correlation colormaps are less correlated and the structure of the polymer is different across different runs. We calculate the Pearson correlation between two colormaps by the following formula. \\
$pc(\alpha \beta)=\dfrac{\langle (p_{ij}^{\alpha}- \langle p_{ij}^{\alpha}\rangle)(p_{ij}^{\beta}-\langle p_{ij}^{\beta}\rangle)\rangle}
{\sqrt{(\langle (p_{ij}^{\alpha}-\langle p_{ij}^{\alpha}\rangle )^2\rangle)((p_{ij}^{\beta}- \langle p_{ij}^{\beta} \rangle)^2 \rangle)}}$ \\

Here $p_{ij}^{\alpha}$ and $p_{ij}^{\beta}$ correspond to the probability of
two segments $i$ and $j$ to be within the cutoff $R_c$ in the colormaps  obtained from the runs (run index $\alpha, \beta$) starting from two different initial conditions. The average $\langle ...\rangle$ has been taken over all values of $i$ and $j$ which have probability $p_{ij}>0.05$ in at least one of the $9$ independent runs.
We chose the pixels with probability $p>0.05$ to avoid the bias from the large parts of dark areas of the colormap, which can result in the high value of $\langle pc \rangle$.    

The average value of the Pearson correlation is plotted in Fig. \ref{fig3} versus the parameter $\sigma$ for the positional colormaps of bacteria {\em C. crescentus} and {\em E. coli}. From the figure \ref{fig3} we see that as we increase the parameter $\sigma$ the value of $\langle pc \rangle$ decreases and the
SD also increases. This suggests that the polymer organizes in different structures in the runs starting from the different initial conditions on increasing 
the value of parameter $\sigma$. Note that even for high values of the parameter $\sigma$ the value of $<pc>$ is nearly $0.5-0.6$ and not lower, this is because the monomers which constitute the CLs will come closer to each other for every value of $\sigma$ as they are connected by the springs. But the rest of the polymer is not able to organize in a particular structure because it cannot overcome the topological constraints.  

\begin{figure}[!hbt]
\includegraphics[width=0.6\columnwidth]{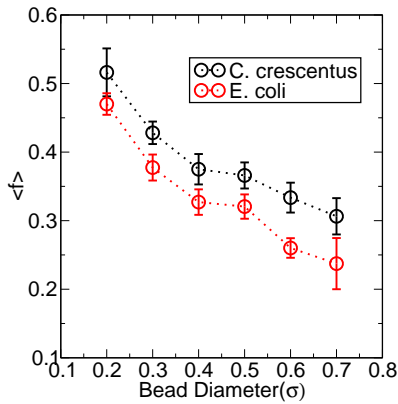}
\caption{\label{fig2}
The y-axis in the figures shows the number of segments $\langle f \rangle$ (normalized by the 
total number of segments) which are at
a distance $<R_c$ with other segments with probability $p>0.05$. The x-axis in the plot shows the parameter $\sigma$.
The average is taken over $9$ independent initial conditions and the standard deviation is
shown by the error bars. 
}
\end{figure}

To quantify the differences  between the colormaps of Fig. \ref{fig16} we calculate $f_i$ the number of segments  which are at a distance $<R_c$  from  segment $i$ with probability $p>0.05$. We can then calculate $ f = \sum_i  f_i/N_{seg}$, normalized by the total number of segments $N_{seg}$, $N_{seg}=80$. 
The value of the cutoff probability $p$ is the same as the value we chose in our previous work \cite{epl,jpcm} and correspond to the deep  red color in the colormap. We then calculate the average value of $\langle f \rangle$
and the standard deviation from $\langle f \rangle$ for $9$ independent initial conditions. In Fig. \ref{fig2}
$\langle f \rangle$ is plotted on the y-axis with the parameter $\sigma$ on the x-axis. From the 
figure, we observe that with the increasing value of $\sigma$, $\langle f \rangle$ is decreasing.
This means less number of segments are found within distance $R_c$ with each other.
This is a non-trivial result because it indicates that the polymer is not only swelling (as was indicated by the previously calculated statistical quantities) but the internal structure of the polymer (as measured by positional correlation colormaps) is also decreasing which in turn  leads to fewer  bright pixels in the colormaps at higher values of $\sigma$. We get the qualitatively similar results for the positional colormaps corresponding to the polymer with the BC-2 set of CLs of bacteria {\em E. coli}. This can be seen in the Fig. \ref{fig2}, where the $\langle pc \rangle$ is plotted corresponding to the colormaps of model chromosome of {\em E. coli}.
  
\vskip1cm

Thus we can conclude from here that the release of the topological constraints are
necessary for a polymer with CLs to organize itself into a particular structure.

\subsection{Study of $\epsilon$}

We now systematically study the role of the parameter $\epsilon$ in the organization of the DNA polymer. As mentioned earlier, for this study we have kept the value of $\sigma$ to be fixed at $0.2a$, and set the cutoff of LJ potential to $r_c= 3\sigma =0.6a$ and vary $\epsilon$ from $0.1 k_BT$ to $0.5 k_BT$.
When we show data for attraction strength $\epsilon=0$, it implies that the $r_c$ is $2^{1/6}\sigma$ with $\epsilon=1$ such that the interaction between the monomers is purely excluded volume. We first calculate the radius of gyration $R_g$ as we vary parameter $\epsilon$ to estimate how the polymer with CLs shrinks in size.  With the increase in the value of $\epsilon$ the value of $R_g$ should decrease because the attraction between the monomers will lead to the collapse of the polymer at higher values of $\epsilon$ which will lead the polymer to form a polymer-globule. 

\begin{figure}[!hbt]
\includegraphics[width=0.6\columnwidth]{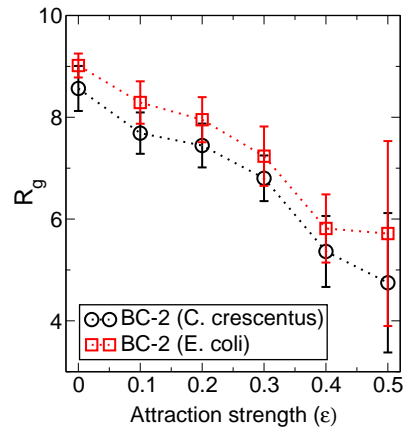}
\caption{\label{fig121}
The plot shows the average value of radius of gyration with the parameter $\epsilon$ for polymer with BC-2 CL-set of bacteria {\em C.crescentus} and {\em E. coli}, respectively. The SD is shown as error bars. The $\epsilon=0$ refers to the case where we cut off LJ at $r_c=2^{1/6} \sigma$, such that only repulsive forces act between non-neighboring spheres.   
}
\end{figure}

The decrease in the value of $R_g$ with increasing $\epsilon$ is shown in the Fig. \ref{fig121}. The error bars  represent the SD from the average value of $R_g$ for the polymer with BC-2 CLs starting from $9$  independent initial conditions. For the higher value of the parameter $\epsilon$ (e.g., $\epsilon=0.5 k_BT$) we see that the errors bars are larger. To understand this, we observe the snapshots of the polymer with CLs starting from the different initial conditions. From the snapshots, we observe that  the bigger loops (which are in the peripheral region for the low value of $\epsilon$) are unable to come on the periphery for higher values of $\epsilon$ in some initial conditions. This is because of the strong attraction between the monomers. It leads to the small value of $R_g$ for some initial conditions and higher for others hence the larger error bar for $\epsilon=0.5k_{B}T$.  

The attraction between the monomers leads to the further compaction of polymer globule in addition to the compaction by CLs. Hence, we expect the number density $n_M(r)$ of monomers to increase in the innermost part as we increase the value of parameter $\epsilon$. But if the packing is too high  for relatively large values $\epsilon$, then the monomers near the center of the globule will be unable to explore the configuration space effectively. This could lead to the polymer getting kinetically stuck. This will also result in very different coarse-grained positional correlation contact maps in different runs. On the other hand, a weak attraction could help get monomers together and increase the positional correlation between different segments 
of the chromosome. We would like  to check if there is an optimum value of  $\epsilon$ which helps the polymer to get compact but at  the same time polymer should be able to explore the configuration space and reach its
organized state. 

\begin{figure}[!hbt]
\includegraphics[width=0.6\columnwidth]{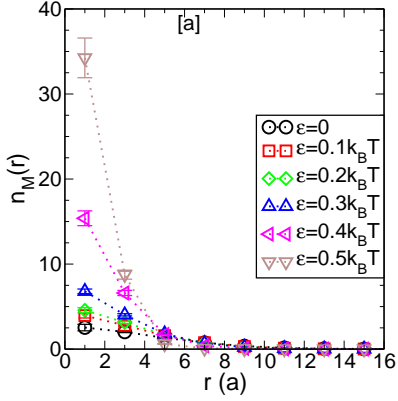}
\caption{\label{fig122}
The figure shows the number density of monomers as a function of distance from the center of mass of the polymer globule. The error bar represent the s.d. from the average value across $9$ independent
initial conditions.
}
\end{figure}

To start,  we plot the number density of monomers with distance from the  CM of the globule in the 
Fig. \ref{fig122}. In the Fig. \ref{fig122} different lines correspond to the different values of parameter $\epsilon$  varying from $0,0.1,0.2.., 0.5 k_BT$. From the figure, we see  that as we increase the value of parameter $\epsilon$ the number density of monomers increases by order of magnitude in the inner core region for $\epsilon=0.4, 0.5 k_BT$ compared to when $\epsilon=0,0.1 k_BT$. But it decays very rapidly with the distance $r$ from the CM of the  globule. This suggests that the inner core is quite dense with monomers for large values of the parameter $\epsilon$ and the monomers in the innermost part of the globule should be less mobile because of the less phase space availability. With this understanding of polymer structure with parameter $\epsilon$ we next investigate the effect of the parameter $\epsilon$ on the internal organization of the polymer.

\begin{figure}[!hbt]
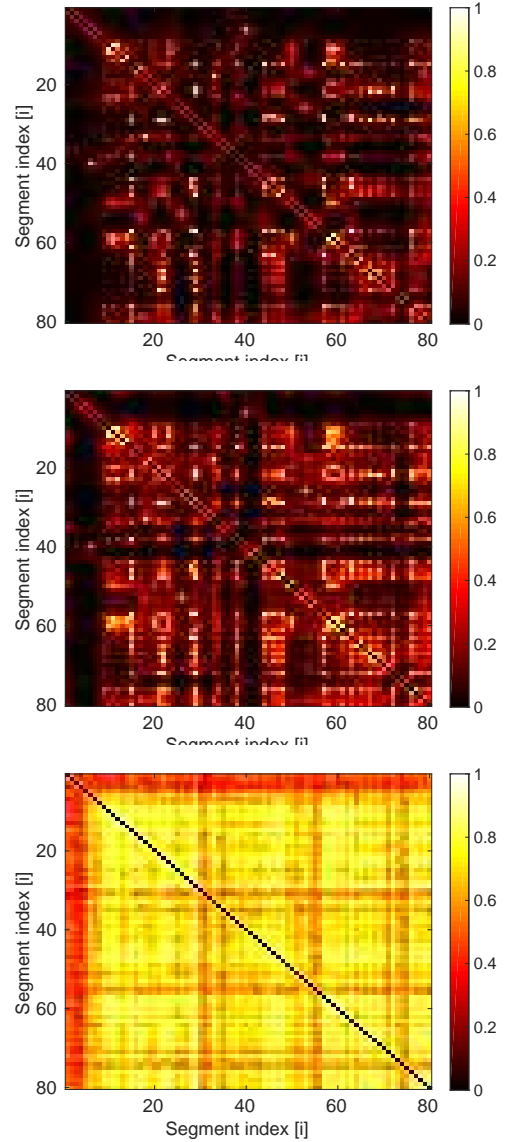

\includegraphics[width=0.75\columnwidth]{caul_cm_dom_corr_exvol02_eps01.eps}
\includegraphics[width=0.75\columnwidth]{caul_cm_dom_corr_exvol02_eps03.eps}
\includegraphics[width=0.75\columnwidth]{caul_cm_dom_corr_exvol02_eps05.eps}
\caption{\label{fig123}
The colormaps show the positional correlation between the center of mass (CM) of
different segments of the polymer for different values of parameter $\epsilon$. There are 80 segments, each with 50 monomers.  The top, middle and bottom figure correspond to the $\epsilon=0.1$,$0.3$,$0.4 k_BT$, respectively.
}
\end{figure}

For this, we calculate the positional correlation of different segments of the polymer as was done
in the previous case. We calculate the positional correlation again by choosing a cutoff $R_c$ such that the value of $R_g/R_c$ remains constant and the same as before. Then we calculate the probability of the CM of the 
two segments to be within a distance of $R_c$. This is an $80\times80$ matrix shown as a colormap in the Figs. \ref{fig123} for three different values of $\epsilon$, 
where the x-axis and y-axis represent the segment index and color represents the probability. 
Bright color corresponds  to the higher probability of the CM of two segments being within disrance $R_c$, and the light color represents the lesser value of probability. From the Figs.\ref{fig123} we notice that as we increase the value of the parameter $\epsilon$ the patch-like pattern in the colormaps become more prominent and clear. But after a certain value of the parameter $\epsilon$ the whole colormap becomes nearly uniformly brighter. This is because the polymer collapses for higher values of $\epsilon$ and each segment is near many other segments with higher probability. The rectangular and square patch-like pattern in the colormap represent that the neighboring segments are coming close to other segments with equal probability, e.g., if segments $i$ and $j$ come within cutoff distance with probability $p_{ij}$ then neighboring segments of $i$ and $j$, i.e., $i+1$,$i+2$ or $j+1$,$j+2$ are also coming within cutoff distance $R_c$ with frequency $\approx p_c$ thus giving the pixels in a rectangular patch of the same color. We do not show the colormaps for the model chromosome of bacteria {\em E. coli} but the outcome and conclusions are qualitatively similar to the colormaps of Fig. \ref{fig123}. 

\begin{figure}[!hbt]
\includegraphics[width=0.6\columnwidth]{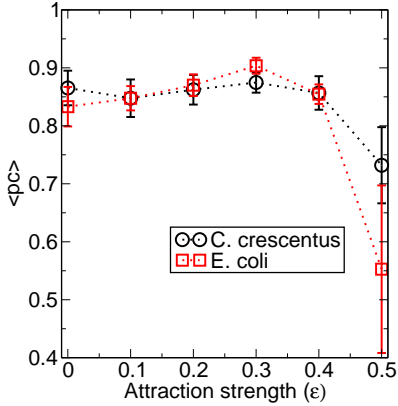}
\caption{\label{fig124}
The plot shows the average value of Pearson correlation for different values of parameter $\epsilon$ for the model chromosome of bacteria {\em C. crescentus} and {\em E. coli}. The SD from the average value is denoted by the error bars. 
}
\end{figure}

If the polymer has a unique organization, then the MC runs starting from independent initial conditions should give statistically similar positional correlation colormaps. To quantify this, we calculate the  average of Pearson correlation $\langle pc \rangle$ of the positional correlations among different runs as was done for the study of polymer organization with various values of the parameter $\sigma$. If the average Pearson correlation 
has a higher value, then we can say that all independent conditions are leading to the same organization
of the polymer. This is shown in the Fig. \ref{fig124} where y-axis and x-axis represent the  value $\langle pc \rangle$ and parameter $\epsilon$, respectively. From the figure we see that the value of $\langle pc \rangle$ increases slightly as we increase the value of the parameter $\epsilon$ till $0.3$ and after that, it decreases. 
This can be because for the larger value of parameter $\epsilon$ polymer from different initial 
conditions may be kinetically stuck in different states and are not able to reach the same structure.
This can also be confirmed from the plots of number density, where for the high value of $\epsilon$
the number density in the inner core is very high. Moreover, we have also tested that after the equilibration runs of $10^6$ MCS, the average distance between the pair of monomers which constitute a CL remains $a$ in all the independent runs. This suggests that the CL monomers are near each other in all independent runs, but other segments of the polymer are not able to organize themselves into a particular structure across different runs starting from independent initial conditions. The plausible reason for this can be the relatively strong attraction between the monomers, because of the strong attraction between the monomers some segments (especially the segments in the inner core) of the polymer are not able to explore the configuration space. Thus giving the different positional correlation colormaps for independent runs. The conclusions from the model chromosome of bacteria {\em E. coli} are similar and can be confirmed from the Fig. \ref{fig124}, where we see that the value of $\langle pc\rangle$ decreases after $\epsilon=0.3k_{B}T$.   

\begin{figure}[!hbt]
\includegraphics[width=0.6\columnwidth]{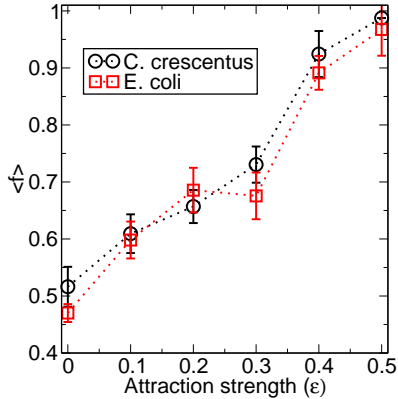}
\caption{\label{fig126}
The plot shows the quantity $<f>$ with the parameter $\epsilon$.
}
\end{figure}

We also quantify the differences in the colormaps of positional correlation using the same quantity
$\langle f \rangle$ which gives the number of pixels with probability $p>0.05$ same as the previous case. This quantity $\langle f \rangle$ is plotted 
in the Fig. \ref{fig126} with parameter $\epsilon$. The average is taken over the colormaps from 
$9$ independent initial conditions. From the graph, we see that the increase in the value of $\langle f \rangle$ 
for lower value of parameter $\epsilon$ is not significant but from $\epsilon=0.3$ to $\epsilon=0.4$ it increases from $0.7$ to $0.9$
and for $\epsilon=0.5$ the value of $\langle f \rangle$ becomes nearly $1$. 
The value of $\langle f \rangle=1$ signifies that all the segments are 
coming closer to other segments. This can also be confirmed from the bottom colormap of positional correlation in Fig. \ref{fig123}, which consists of all the bright pixels. We obtain the same conclusions for the colormaps of the polymer with BC-2 set of CLs of {\em E. coli}, see Fig. \ref{fig126}. 

So from these results, we can say that there is an optimum value of attraction strength between the monomers which helps the polymer to organize. Since for smaller attraction strengths $\epsilon$, the polymer will be able to explore configuration space and can obtain a particular organization but for the very high value of attraction strengths the polymer can get kinetically stuck into different states and would not be able to explore the configuration space.

\section{Discussions}
We study the effect of the release of topological constraints and molecular crowders in the organization of the polymer with very few  cross-links ($\approx 3\%$ of the monomers) taken from the experimental contact map of bacteria {\em E. coli} and {\em C. crescentus}. We showed that the release of topological constraints is crucial for the polymer to organize into a particular structure since the model polymer from $9$ independent runs are able to organize into a particular structure when we allow the chain to cross itself and release the topological constraints. Thus, we think that the activity of enzyme topoisomerase can play a vital role in the organization of the chromosome at large-length scales by allowing the chains to cross itself. We  are unable to comment on the frequency of the activity of enzyme topoisomerase in in-vivo cells to promote chain crossing but compare with the frequency of chain crossing in our simulation but show that the release of topological constraints is necessary for a polymer to get organized in a unique structure.  We also show that
the molecular crowders in the bacterial cytoplasm of the cell calead to an optimum attraction between the DNA segments so that the DNA polymer is able to explore the different configurations as well as should be able to organize into a particular structure. We find the value of effective attraction between the monomers to be $\epsilon=0.3k_{B}T$ for which the polymer from different initial conditions is able to obtain a particular organization. 

We want to study the role of individual factors in the organization of the bacterial chromosome, thus we do not introduce the effect of cell wall confinement in the present paper but, we study the effect of confinement with the effect of these factors by taking the optimum values of the parameter from this study and report it in the separate accompanying paper part -II.

\section{Acknowledgements}
We acknowledge the use of computer cluster bought from DST-SERB Grant No.  EMR/2015/000018  and  funding  from  DBT  Grant BT/PR16542/BID/7/654/2016 to A. Chatterji. A.C. acknowledges funding support by DST Nanomission, India under the Thematic Unit Program (Grant No. SR/NM/TP-13/2016). 
\bibliographystyle{apsrev4-1}
\bibliography{paper3}
\clearpage

\end{document}